\documentclass[a4paper,fleqn]{cas-dc}

\usepackage[authoryear,longnamesfirst]{natbib}
\usepackage{color,soul}
\usepackage{listings}

\def\tsc#1{\csdef{#1}{\textsc{\lowercase{#1}}\xspace}}
\tsc{WGM}
\tsc{QE}

\begin{document}
\let\WriteBookmarks\relax
\def\floatpagepagefraction{1}
\def\textpagefraction{.001}

\shorttitle{PLC-VBS, a PLC Control Logic Vulnerability Scanning Tool}

\shortauthors{S Maesschalck et~al.}

\title [mode = title]{Walking Under the Ladder Logic: PLC-VBS, a PLC Control Logic Vulnerability Scanning Tool}    



%

\author[1]{Sam Maesschalck}[auid=000,bioid=1,
                        orcid=0000-0003-4609-3487]

\cormark[1]


\ead{s.maesschalck@lancaster.ac.uk}

\credit{Conceptualization, Methodology, Validation, Investigation, Data Curation, Writing - Original Draft, Writing - Review \& Editing, Visualization}

\affiliation[1]{
    organization={Lancaster University},
    addressline={InfoLab21}, 
    city={Lancaster},
    postcode={LA1 4WA}, 
    country={United Kingdom}}

\author[1]{Alexander Staves}[auid=000,bioid=2,
                        orcid=0000-0002-9861-0477]
\ead{a.staves@lancaster.ac.uk}
\credit{Conceptualization, Methodology, Investigation, Writing - Original Draft, Writing - Review \& Editing}

\author[1,2]{Richard Derbyshire}[auid=000,bioid=3,
                        orcid=0000-0003-3902-5056]
\ead{ric.derbyshire@orangecyberdefense.com}

\credit{Conceptualization, Methodology, Writing - Original Draft}

\affiliation[2]{organization={Orange Cyberdefense},
    addressline={250 Waterloo Road, South Bank Third Floor}, 
    city={London},
    postcode={SE1 8RD}, 
    country={United Kingdom}}

\author[1]{Benjamin Green}[auid=000,bioid=4]
\ead{b.green2@lancaster.ac.uk}
\credit{Conceptualization, Methodology, Investigation, Software, Writing - Original Draft, Visualization, Writing - Review \& Editing, Supervision}

\author[1]{David Hutchison}[auid=000,bioid=5,
                        orcid=0000-0001-6052-0559]
\ead{d.hutchison@lancaster.ac.uk}
\credit{Conceptualization, Methodology, Writing - Review \& Editing, Supervision}

\cortext[cor1]{Corresponding author}



\begin{abstract}
Cyber security risk assessments provide a crucial starting point towards the understanding of existing risk exposure, via which suitable mitigation strategies can be formed. Risk is viewed as a product of threat, vulnerability and impact, and equal understanding of each of these elements is vitally important. This can be a challenge in Industrial Control System (ICS) environments, where adopted technologies are typically not only bespoke, but interact directly with the physical world. To date, existing vulnerability identification has focused on traditional vulnerability categories. While this approach provides risk assessors with a baseline understanding and the ability to hypothesize about potential resulting impacts, it is rather high level, operating at a level of abstraction that would be viewed as incomplete within a traditional information system context. The work presented in this paper takes the understanding of ICS device vulnerabilities a step deeper. It offers a tool, PLC-VBS, that helps identify Programmable Logic Controller (PLC) vulnerabilities,  specifically within logic used to monitor, control, and automate operational processes. PLC-VBS gives risk assessors a more coherent picture about the potential impact should the identified vulnerabilities be exploited; this applies specifically to operational process elements.
\end{abstract}


\begin{highlights}
\item A previously unconsidered approach towards the understanding of PLC vulnerabilities.
\item A tool, PLC-VBS, supporting the identification of vulnerable PLC memory.
\item The results of a vulnerability scan conducted against a vendor-provided PLC code base commonly used in practice.
\end{highlights}


\maketitle

\section{Introduction}
\label{introduction}
\begin{figure}
	\begin{center}
    	\includegraphics[width=1\columnwidth]{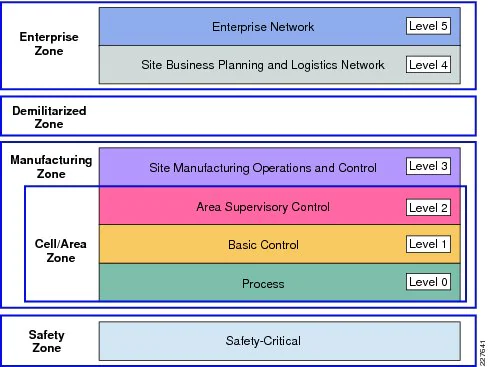}
	\end{center}
    \caption{Purdue Model~\cite{Didier2011}}
    \label{fig:purdue}
    \vspace{-15pt}
\end{figure}

Critical National Infrastructure (CNI) can be seen as the supporting backbone of everyday life, typically categorised across 13 sectors, such as energy and water~\citep{CPNI2020}. Underpinning the core operational elements of these environments are Industrial Control Systems (ICSs). The Purdue Enterprise Reference Architecture (Figure~\ref{fig:purdue}) provides a clear depiction of an ICS split into hierarchical levels. From this model a defined set of operational functions is outlined, each of which plays its role in the uninterrupted operation of physical operational processes (treatment of water, distribution of electricity, etc.), of paramount importance to societal well-being. The lower the level within this model, the closer to physical operational processes the associated devices become, with an increasing deviation from traditional IT towards more bespoke industrial technologies.

Over recent years we have observed an increasing level of sophistication in attacks targeting ICSs that underpin CNI~\citep{miller2021}. While legislation has been introduced to support the continued development of defensive strategies~\citep{EuropeanCommission2019}, there remains a challenge using available resources to accurately identify and assess risk. Where risk is viewed as the product of threat, vulnerability, and impact, understanding each element is of equal importance~\citep{cost}. Vulnerability and impact can present a significant challenge to be fully comprehended within an ICS context, as current tooling focuses on traditional vulnerability categorisation at a device level (Denial of Service, protocol limitations, etc.). This limits an assessor's ability to focus on how an attacker could impact operational process behaviour, which is of particular importance when one considers Advanced Persistent Threats~\citep{ahmad2019}.

The work presented in this paper takes the understanding of ICS device vulnerabilities a step further, and offers a tool that we call PLC-VBS (Programmable Logic Controller Variable Block Scanner) that helps identify Programmable Logic Controller (PLC) vulnerabilities. More specifically, it can identify vulnerabilities within control logic (program code) used to monitor, control, and automate operational processes (e.g., a water tank level). The insights it affords to risk assessors also allow for a more coherent picture on the potential impacts should the identified vulnerabilities be exploited, with direct links to specific operational process elements (e.g., a valve within a water tank level control process). We present the results of a practical usecase study with commercial-grade PLCs and publicly available control logic, demonstrating the value of our approach and how it can enhance risk assessment processes when compared to traditional vulnerability scanning tools. In this research we have focused on vendor-provided library functions as these are widely used in practice~\citep{Ljungkrantz2007}.

The primary contributions of this paper are:

\begin{itemize}
	\itemsep0em
	\item A previously unconsidered approach towards the understanding of PLC vulnerabilities.
	\item A tool, PLC-VBS, supporting the identification of vulnerable PLC memory.
	\item The results of a vulnerability scan conducted against a vendor-provided PLC code base commonly used in practice.
\end{itemize}

The remainder of this paper is structured as follows. Section~\ref{relatedwork} covers related work. Section~\ref{PLCbackground} provides a background on PLC program structures. Section~\ref{scannertheory} develops our main contribution, which is subsequently validated in Section~\ref{PoC}. Section~\ref{discussion} provides a discussion of the results, with Section~\ref{conclusion} concluding the paper and suggesting areas for future work.

\section{Industrial Control Systems}
\label{background}

When looking at national infrastructure, we can see that critical parts of it are underpinned by industrial control systems. These systems are designed to control and automate industrial processes used within many industries, including water and nuclear~\citep{Green2017, McLaughlin2016}. Critical infrastructure consists of 13 national infrastructure sectors~\citep{CPNI2021}: Chemicals, Civil Nuclear, Communications, Defence, Emergency Services, Energy, Finance, Food, Government, Health, Space, Transport and Water. Of these sectors, ICSs can be found in nearly all of them, for example, as part of manufacturing plants to control robots or the health sector to control HVAC (Heating, Ventilation, and Air Conditioning) systems.

Any adverse impact on the operation of ICSs within organisations that host critical infrastructure can severely impact society. Therefore, any vulnerabilities that can be found within these systems pose a potential danger. Over the past decade, we can see a trend of Internet-connected ICSs, expanding the attack surface on these devices as it would allow adversaries from anywhere on the Internet to interact with these systems. Many ICSs were generally not designed with this Internet connectivity in mind, and the proprietary protocols they use may be susceptible to attack~\citep{Ahmed2017}; also, the software running on these systems is often not regularly patched and can stay vulnerable for a long time.

This paper mainly focuses on PLCs~\citep{erickson1996}, which are used to control physical devices such as valves and actuators. In addition to this, they provide management of processes within the environment and monitor the status of the system. PLC logic can, for example, be used to trigger an alarm when an event (e.g. water level rises to 90\%) happens. The logic of the PLC is written in ladder logic. These systems are often paired with a Human Machine Interface (HMI)~\citep{CISA-HMI} which allows an operator to control the physical device using a graphical user interface, rather than having to send a specific command to the PLC as the HMI sends this command instead based on the input it has been given. The way this is done is by directly interacting with the PLC memory structure and changing values related to the operation that is requested, e.g. changing the valve from closed (0) to open (1). An HMI can also be used to display sensor information in a graphical manner, such as the water level within a tank or historical data from within the environment.

\section{Related Work}
\label{relatedwork}
Research into vulnerabilities within traditional IT systems/software has been at the forefront of security research for several decades. This is in direct contrast with work in ICSs, which has seen an increased focus only in recent years. This aligns with the shift towards more integrated IT and ICS environments, removing the isolation between the two, rendering ICS environments more vulnerable to external adversaries, including those from nation states~\citep{Derbyshire2018,miller2021}. Due to ICS deployment issues, there exists a consequent plethora of vulnerabilities. However, research within the ICS vulnerability space has tended to focus on operating systems, firmware, industrial protocols and circumventing traditional security measures~\citep{abbasi2016ghost, Nochvay2019, drias2015taxonomy, wardak2016plc, Biham2019}. This includes research and development into vulnerability scanning tools designed with ICSs in mind~\citep{antrobus2016, antrobus2019}. These tools are critical in providing information on exploitable vulnerabilities within operational environments during risk assessments.

Two notable vulnerabilty scanners have been developed by~\cite{antrobus2016, antrobus2019}. One, SimaticScan, focuses on vulnerability scanning of Siemens PLCs, with the other, PIVoT Scan, focusing on scanning Industrial Internet of Things (IIoT) environments. SimaticScan is designed to detect a range of vulnerabilities within a Siemens PLC, including Web vulnerabilities and unauthorised read/write requests. Within its evaluation, this tool identified multiple vulnerabilities across the researchers' testbed environment, including one that was previously unknown. In the evaluation of PIvoT Scan it was seen to outperform a popular vulnerability scanner, Nessus. Although Nessus \citep{nessus} does contain plugins that support the scanning of ICS devices, it remains a tool primarily developed for use within traditional IT environments. OpenVAS \citep{openvas}, another well known vulnerability scanner, lacks ICS specific scanning capabilitites, limiting its value in a similar way to Nessus. While both OpenVAS and Nessus are able to find basic vulnerabilities (e.g., DoS, HTTP-based, and SMTP-based)~\citep{mcmahon2018}, they fail to identify vulnerabilities directly linked to operational process monitoring, control, and automation. Two additional ICS specific port scanners that can be used to assess vulnerabilities are PLCScan~\citep{PLCScan} and ModScan~\citep{Bristow2008}. These tools are not vulnerability scanners; however, their output (software and hardware focused) can be used in conjunction with vulnerability databases to identify the existence of known vulnerabilities. Therefore, they can not be used in the identification of new vulnerabilities that might affect the system under consideration.

PLC programs are developed across five special programming languages (see Section~\ref{PLCbackground}), one of which is commonly referred to as "ladder logic" (ladder diagrams). Ladder logic was originally designed to document the design and construction of relay racks in a sequential process. Over time, ladder logic evolved into a full programming language, and with it, related research. Research on ladder logic by~\cite{kottler2017formal} introduces formal verification, where~\cite{eckhart2019security} allow us to consider wider implications of vulnerabilities within systems through the use of a security development lifecycle. As with conventional programs, adversaries could identify and exploit issues within a PLCs program. Therefore, the ability to identify and assess vulnerabilities within programs running on these systems is of crucial importance.

When evaluating PLC programs, specific programming practices have to be taken into account~\citep{Fluchs2020}. Current programming practice includes the use of library functions developed and provided by device vendors. This can be seen in~\cite{Ljungkrantz2007}, in which the authors explore programming practices and the use of library functions. We can assert that these functions, and consequently associated memory use, behave in the same way across all devices in which they are deployed. Additionally, not only do PLC programmers rely on these library functions, but they also create their own custom functions, written for use across an entire organisation. Identifying vulnerabilities within these functions would render all systems in which they are deployed vulnerable. Vulnerabilities on this scale would be very considerable, given the international breath of their deployment.

Tools such as PCaaD~\citep{green2021} (Process Comprehension at a Distance) take an initial step towards the identification of vulnerabilities within PLC programs and allow for the development of process comprehension~\citep{Green2017a} without any pre-existing knowledge about the target system; however, they are highly focused and fail to assess previously unseen PLC programs due to their use of signature-based recognition. This, along with the other work discussed throughout this section, provides the primary motivation for the development of a PLC vulnerability scanning tool capable of holistically identifying issues in PLC programs, whether created by vendors as part of public libraries, or developed in-house by PLC programmers.

Thus, we propose such a new tool, Programmable Logic Controller Variable Block Scanner (PLC-VBS), which provides a comprehensive overview of the vulnerabilities within the control logic of the PLC. Due to the focus on ladder logic, our tool is vendor agnostic and specifically designed for PLCs rather than re-purposed from IT-based tools. With this tool, programming practices can be improved and evaluated to create a more secure system, as will be explained in the subsequent sections of this paper.

\section{Scanning for Vulnerabilities}
\label{scannertheory}
Based on the already mentioned POUs, we have two options when it comes to the development of a scanner capable of identifying control logic vulnerabilities; (1) focus directly on the control logic itself, or (2) examine how PLC memory is being used within the control logic to store variable data (gVBs and fVBs). The following subsection will introduce a threat model, defining vulnerabilities a scanner should seek to identify. This will be used to ground discussions on the two alternative scanner approaches.

\subsection{Threat Model}
\begin{figure}
	\begin{center}
    	\includegraphics[width=1\columnwidth]{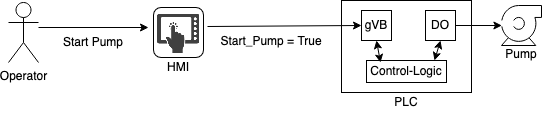}
	\end{center}
    \caption{HMI - PLC Interaction Process}
    \label{fig:process}
\end{figure}

Existing work has demonstrated that attacks targeting PLCs can be executed at both the network and hardware level~\citep{Green2017a}. These have been witnessed across a range of real-world events dating back to the 1980s~\citep{miller2021}. One of the most commonly discussed attacks, operating at the network level, exploits the use of PLC network protocols to inject unauthorised command messages into a PLC's memory (VBs)~\citep{MITRE2021, Robles-Durazno2019}. Figure~\ref{fig:process} depicts normal/trusted operational use of a Human Machine Interface (HMI) by a system operator. In simple terms, an operator will use HMIs (these can be compact touch screen devices~\citep{SiemensHMI}, or larger desktop/server based systems~\citep{Siemenswincc}) to send requests to PLCs. These requests are stored in the PLCs VBs (gVBs and/or fVBs), which in turn are ingested and processed as part of the PLC control logic. The control logic will then operate underlying physical processes (e.g., starting/stopping a pump) via output cards (See the Digital Output "DO" card in Figure~\ref{fig:process}). It is this trusted HMI functionality that attackers mimic when adopting unauthorised command message attack techniques.

As this type of attack can be challenging to mitigate at the network layer~\citep{green2021}, it offers a compelling argument towards identification and remediation at the device level (i.e., within the control logic), and therefore forms the basis for vulnerability identification through the remainder of this paper.

\begin{figure}
	\begin{center}
    	\includegraphics[width=1\columnwidth]{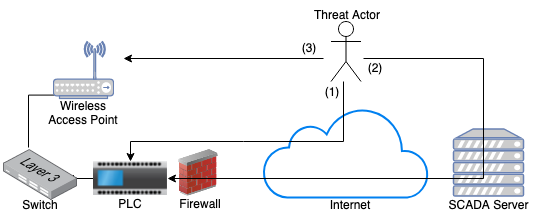}
	\end{center}
    \caption{Threat Model}
    \label{fig:threatmodel}
\end{figure}

Attacking a PLC via the injection of malicious data into its memory requires network-level access via the device's supported network protocol and function (e.g., S7-Comm and Write requests~\citep{Molenaar2013}). This can be achieved in a number of ways, with Figure~\ref{fig:threatmodel} included as a point of reference across the following examples:

\begin{itemize}
    \item (1) The PLC has been configured with a public IP address. Examples of this type of deployment can be seen in Shodan~\citep{Shodan2020}. As the PLC resides directly on a public IP address with no network-based defensive controls, an attacker with Internet access can directly inject malicious data into the PLC's memory with no restrictions.
    \item (2) The PLC has been configured to sit behind a firewall that filters incoming requests. In this scenario, an attacker would be required to compromise a trusted device that is permitted to traverse the firewall and communicate with the PLC.
    \item (3) The PLC is within a network segment that has a poorly protected WiFi network. Once access has been obtained, an attacker would have direct network connectivity to the PLC.
\end{itemize}

\subsection{Scanning control logic}
\label{scanning_cl}
Figure~\ref{fig:ladder} provides an example of a Ladder Diagram, in which an output (Example\_Output) is controlled via the use of two inputs (Example\_Input\_1 and Example\_Input\_2). A vulnerability scanner capable of identifying vulnerabilities within control logic would be required to interpret this, and flag the presence of known vulnerabilities based on a set of pre-defined signatures. For example, using the control logic from Figure~\ref{fig:ladder} as a base for discussion, what state should "Example\_Output" be when "Example\_Input\_1" and "Example\_Input\_2" are in opposing states (i.e., one is in a 1 state and the other a 0)? There is no control logic to manage this state combination; therefore, this would induce a vulnerability whereby an attacker is able to manipulate "Example\_Output" via unauthorised command messages (during time periods where the Example\_Inputs are in opposing states). This type of vulnerability would need to be included as part of a pre-defined set of signatures.

A vulnerability scanner adopting this approach would be expected to meet the following set of requirements:

\begin{itemize}
    \item Support all five PLC programming languages as defined by BSI/IEC 61131-3:2013~\citep{BritishStandardsInstitute2013}, and any subtle vendor-specific offshoots.
    \item Develop and maintain a database of PLC programing practices that can induce vulnerabilities.
    \item Allow for the extraction of control logic directly from a PLC or via a project file.
\end{itemize}

These requirements present several challenges. As vendors typically use their own custom PLC programming software~\citep{Siemens2020}, they will compile control logic for use on their devices alone. Therefore, when PLC code is extracted from a PLC, normalisation of data will be required to provide a standardised output for which vulnerabilities can then be identified. In addition, the development and maintenance of a comprehensive vulnerability database will require significant upkeep, and must be compatible with all five PLC programming languages.

Despite these challenges, there are clear benefits in adopting this approach. Primarily, upon the identification of a vulnerability, it will be possible to provide focused and proportionate guidance on its impact, and how it can be remediated.

\subsection{Scanning PLC Memory}
\label{scanning_mem}
As previously noted, PLCs provide direct access to memory over the network as part of their core operational functionality. This forms the basis of our threat model. Therefore, examining the construction and scope for manipulation of PLC memory, in a similar way to that of an attacker, provides an alternative approach to vulnerability identification.

Taking the example from Section~\ref{scanning_cl}, whereby the manipulation of "Example\_Output" in Figure~\ref{fig:ladder} is made possible during time periods in which the "Example\_Inputs" are in opposing states, a tool scanning PLC memory would be required to identify this vulnerability through the direct manipulation of "Example\_Output" during a scan.

A vulnerability scanner adopting this approach would be expected to meet the following set of requirements:

\begin{itemize}
    \item Support for all major PLC communication protocols that provide direct access to memory.
    \item Identification of memory that can be directly manipulated as well as indirectly (i.e., can "example\_output" be directly manipulated or can it only be manipulated via the two "example\_inputs"?).
    \item Account for control logic overwriting manipulated memory with correct/true data within a given time-period.
\end{itemize}

The primary challenge in meeting these requirements comes in the form of protocol support. As existing protocols vary in their associated open-source documentation, practical manual dissection of their structure may first be required. In addition, unlike direct scanning pf control logic, even where all requirements are met, a scanner adopting this approach will still rely on manual examination of the control logic to better understand why areas of memory are open to manipulation, and how appropriate remediation activities can then be undertaken.

\subsection{Summary}
Section~\ref{relatedwork} identified various existing works and tooling, developed for the identification of PLC vulnerabilities. However, while these approaches adopt techniques that may be more suitable for use in an industrial, safety-driven context, they still focus on traditional sorts of vulnerabilities. While there are exceptions, they still fail to explore control logic induced vulnerabilities, instead offering other related features, such as code verification and process comprehension.

In defining a threat model, we have focused the functionality of a control logic vulnerability scanner down to the use of a specific attack vector, namely unauthorised command messages injecting malicious data into PLC memory~\cite{MITRE2021}.

By reviewing the two scanning techniques presented here, we believe scanning PLC memory offers a lower barrier to entry in the identification of control logic vulnerabilities, and affords a greater level of scalability. This is due to the commonality of memory use/access across PLCs from multiple vendors~\citep{green2021}, alongside commonly adopted network protocols~\citep{feld2004profinet, tamboli2015implementation}. A tool developed on one device could be used directly on another with no requirement for further adaptation. Furthermore, as this approach operates in a similar way to the attack vector defined within our threat model, the scanner could be executed from specific network segments, thus adding to the users' overall understanding of risk exposure where an attacker has an assumed level of system access. This would allow for the simultaneous evaluation of control logic and access controls.

\section{Scanner Proof of Concept}
\label{PoC}
To facilitate practical proof-of-concept exploration of our scanner a Siemens 300 series PLC was used based on its global adoption (making it a representative use-case)~\citep{Siemens2020a}, and the Siemens TIA v13 (Professional) platform as a programming agent~\citep{Siemens2020}. All evaluation was undertaken within a physical environment deployed following the guidance provided by~\citep{green2020}.

\subsection{Siemens PLC Ecosystem}
\label{SiemensEco}
The Siemens 300 series PLC used during our evaluation supports Ladder Diagrams, Statement List (Instruction List), Function Block Diagram, Graph, and Structured Text programming languages. During programming, four primary blocks are used to build the control logic: Organisation Blocks (OB), Function Blocks (FB), Functions (FC), and Data Blocks (DB) (i.e., Variable Block POUs). These four blocks map directly with the POUs defined within Section~\ref{PLCbackground}.

Within these blocks, aside from DBs, an engineer can write control logic. As DBs in the Siemens Ecosystem represent VBs, they are used to store variables called OBs, FCs, and FBs. There are additional symbol types where data can be generated, output, and stored, such as I/O Signals (I and Q), Marker Memory (M, MB, etc.), Peripheral I/O (PIW, PQB, etc.), and Timers and Counters (T \& C)~\citep{PLCDev2020}. However, for the purposes of this work we will be focusing on DBs as these store the variables related to the linked function which we seek to evaluate.

\begin{figure*}
    \centering
    \includegraphics[width=\linewidth]{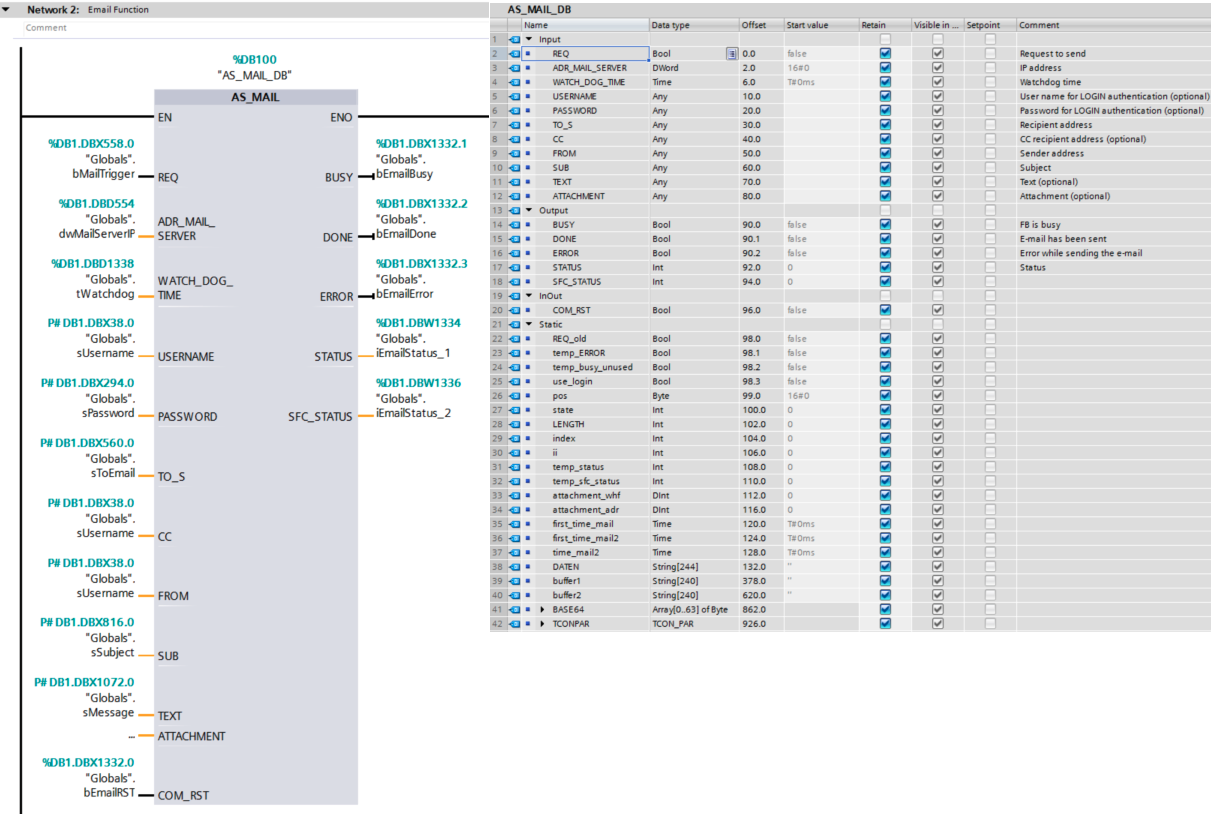}
    \caption{Email Function and Associated VB (DB 100)}
    \label{fig:emailfunction}
\end{figure*}

Figure~\ref{fig:emailfunction} has been included to better explain the use of gVBs and fVBs within a Siemens context. To the left of this figure is a single rung of ladder logic. On this rung is the AS\_MAIL FB, part of the Siemens TIA Communications library. Once triggered, this FB will send a pre-constructed email to a pre-defined recipient. This could be used as part of an emergency alerting system, notifying system operators of issues with the underlying operational process (e.g., a pump tripping). To the right of this figure is the AS\_MAIL FB's associated fVB (DB100), where data for all of its local variables are stored. The left hand side of AS\_MAIL in Figure~\ref{fig:emailfunction} depicts a set of inputs, required in order for the FB to operate. These inputs can be specified using four techniques: 

\begin{itemize}
    \item gVBs - As can be seen in "REQ", where the value stored in "bMailTrigger" (gVB address DB1.DBX558.0) is copied/pasted into "REQ" (fVB address DB100.DBX0.0) during every control logic cycle.
    \item Pointers - As can be seen in "USERNAME", where the gVB address of "sUsername" (P\#DB1.DBX38.0) is copied and pasted into "USERNAME" (starting at fVB address DB100.DBX10.0) during every control logic cycle. Here the value is simply read from the gVB via the pointer address on every cycle, rather than being copied/pasted from the gVB into the fVB.
    \item Defaults - As can be seen in "WATCH\_DOG\_TIME" (T\#0ms in a grey font). Here, instead of specifying an input value, the fVBs default value of 0ms has been left in place.
    \item Direct Inputs - Figure~\ref{fig:emailfunction} does not contain an example of this input type; however, it would look the same as a default input, but with a blue font. Here the value is input directly within the ladder logic, and would be copied/pasted into "WATCH\_DOG\_TIME" (fVB Address DB100.DBD6) on every cpu-cycle, in the same way as if configured via a gVB.
\end{itemize}

As previously noted, PLC-VBS will focus specifically on VBs (i.e., DBs in the Siemens ecosystem), and the ability to manipulate data stored within them via unauthorised command message attacks. From the example provided in Figure~\ref{fig:emailfunction}, this would include DB1 and DB100 VBs.

\subsection{Vulnerability Scanner}
\label{practicalSiemens}

\begin{figure}
    \centering
    \includegraphics[width=1\columnwidth]{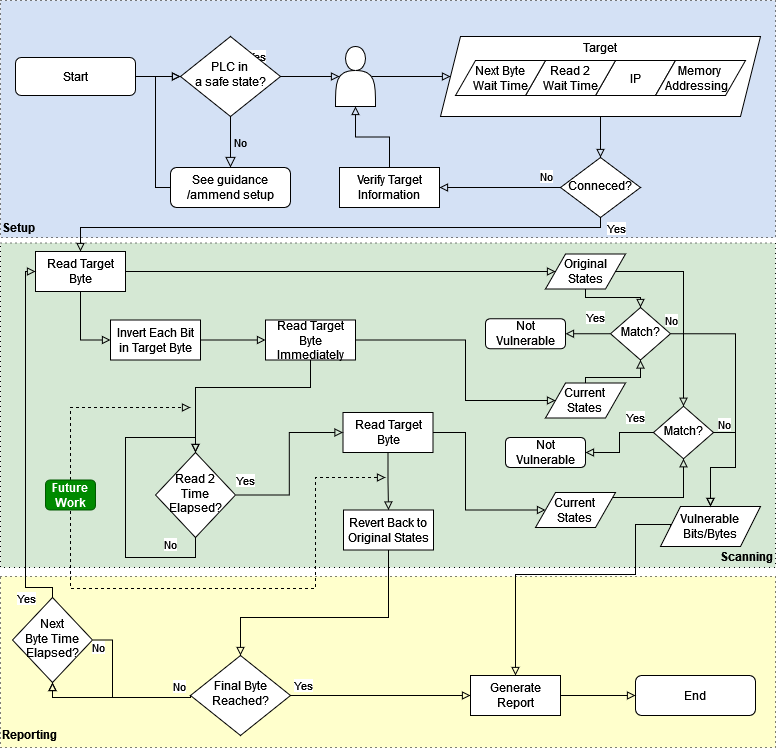}
    \caption{Vulnerability Scanner Process}
    \label{fig:tool}
\end{figure}

Within this subsection we first provide a high-level overview of PLC-VBS, before deconstructing it into multiple phases where we provide further in-depth detail about its functionality. Figure~\ref{fig:tool} provides an overview of the scanner's operation, as a high-level reference point.

When starting the scanner, the user is required to provide the PLC's IP address, the location of its CPU (rack and slot numbers), target VB (DB number), the time period between writing a new test value to a VB and then reading it back (\textbf{Read 2 Time}), and the wait-time between each byte being tested (\textbf{Next Byte Time}). Also, before starting the scanner, the user must first confirm the system is in a safe state. This typically involves removing any physical outputs to operational equipment (e.g., pumps and valves) and other devices (e.g., PLCs and HMIs), that could enact operational change based on its actions. Once a connection to the PLC is established, the scanner begins testing each byte sequentially within the VB. The scanner uses S7Comm read and write requests, the same requests used by engineers and HMIs to communicate with Siemens PLCs~\citep{green2021}. For each byte it scans, the scanner attempts to invert each bit in the target byte, before reading it back to confirm if the inversion request has been accepted and retained by the PLC. If the states of bits within the byte have changed, they, and the byte itself, are classified as vulnerable. The scanner then waits until the time specified by the user has elapsed (\textbf{Read 2 Time}), before reading the byte again. If the bits have remained in the inverted state, we know the inversion request has persistence (i.e., you only need to send one write request to change the state of the bits). If the bits have reverted back to their original state, they have been overridden by the control logic, and therefore any attempts to change their state will not be persistent with a single write request. Finally, we revert the byte back to its original state and wait until the defined time period has elapsed (\textbf{Next Byte Time}) to begin testing the next byte. When the final byte has been reached a report is shown and the scanning ends. 

\subsubsection{\textbf{Scanner Phase 1 (Setup)}}
\label{scanphase1}
involves evaluating the current PLC state and preparing the scanner for use. Ensuring the PLC is in a safe state is important to reduce the risk of inadvertent operational impact. Once verified, the user can continue with the scanner's baseline setup, which determines what the scanner will target and how it will operate. Aside from the target PLC's IP address and CPU details, two timers must be configured. The first is a wait-time (\textbf{Read 2 Time}) between inverting each bit within a byte and then reading it back (to check for inversion persistence); second (\textbf{Next Byte Time}) is the wait-time between each byte under test (only a single byte is tested at any one time). Checking for persistence is vital, as some bits will be overridden by other control logic operations, which would then require a consistent flow of write requests in an attempt to achieve the desired level of persistence (an unreliable approach which may be more noticeable on network monitoring solutions). The wait-time between each byte under test is used to reduce load on the PLC, and setting this to 0 could impact the results as the PLC cannot respond in a timely manner or may fail to process scanner requests. We recommend this to be at least 1 second. Finally, the user is required to specify the VB under review, thus constraining the scanner to an approved area of PLC memory. Once the user has completed these actions, the scanner attempts to connect to the PLC, should the connection fail, the user is asked to verify the information provided.

\subsubsection{\textbf{Scanner Phase 2 (Scanning)}}
\label{scanphase2}
begins once the scanner's initial setup has been completed and a connection to the PLC has been established. Within this phase, the scanner's main objective is to iterate through each byte within the memory space defined during its setup. The scanner starts by reading the first byte within the memory space (S7Comm read request) and stores its state (the byte's \textbf{original state}). Then, each bit within the target byte is inverted (S7Comm write request) and the byte is read again. The \textbf{current state} of the byte is stored, and used as a comparison against the original state. If these states match, there are no vulnerable bits within this byte, and the byte as a whole cannot be manipulated. Where the states do not match, the byte is identified as vulnerable. The scanner then logs both the vulnerable byte and the individual vulnerable bits. 

Once the first wait-time (\textbf{Read 2 Time}) has elapsed, the scanner reads the target byte for a third time and stores its current state. Then, as before, the current state is compared with the original state. The results of this will determine whether or not the inversion has persistence, and if so, across which bits. The scanner logs both the vulnerable byte and the individual bits. Finally, the byte is reverted back to its original state (S7Comm write request).

\subsubsection{\textbf{Scanner Phase 3 (Reporting)}}
\label{scanphase3}
has two functions. If the final byte of target memory has not yet been reached, the scanner waits until the second wait-time (\textbf{Next Byte Time}) has elapsed. It then returns the scanner back to phase two where it begins testing the next byte. Once the final byte has been reached, the tool generates a report identifying the vulnerable bits and bytes within the scanner memory space. The tool then exits.

\subsection{Experimentation}
\subsubsection{Method}
In order to evaluate the effectiveness of PLC-VBS we require access to real-world control logic. This presents a significant challenge, particularly when seeking to identify vulnerabilities, as an organisation sharing internally developed control logic would be unlikely to provide approval for its discussion within the public domain. Therefore, we opted to use Library FBs written by Siemens, included as part of their PLC programming agent TIA Portal~\citep{Siemens2019}. As per our findings from Section~\ref{relatedwork}, vendor provided control logic is extremely common, is context agnostic (it can be applied to water treatment, energy distribution, manufacturing processes, etc.), and is internationally adopted, making it an ideal candidate for exploration.

We selected ten FBs, offering a range of functionality, from basic counts, to more complex data exchanges. Each FB has its own associated fVB, as per the example shown in Figure~\ref{fig:emailfunction} and discussions across Section~\ref{SiemensEco}. Furthermore, these fVBs are accessible over the network, so are directly applicable to the approach adopted by our scanner. Where possible, each FB was configured multiple times, adopting three of the four configuration input techniques described in Section~\ref{SiemensEco}:

\begin{itemize}
    \item gVBs
    \item Defaults
    \item Direct Inputs
\end{itemize}

As each FB dictates the type of input it is willing to accept, it was only possible to use pointers where they were specifically requested (hence their exclusion here). This means if a FB requests a pointer as an input, no other input type can be applied, preventing us from assessing alternatives. Some fVB variables did not include default values, so where possible we left these empty (optional inputs), and some fVB variables did not allow for direct inputs. Where direct inputs were not permitted, we used gVBs. We ran our tool once per configuration setup, and logged the findings before moving on to the next FB.

\subsubsection{Results}
\label{results}
We ran our scanner across ten fVBs, see Table~\ref{tab:examplefunctions} in the appendix for a brief summary of each FB.
As per our method, each FB was configured three times, with associated results shown in Tables~\ref{table:functions_experiment} and~\ref{table:functions_experiment_delay} (i.e., defaults, direct inputs, and gVBs).

High-level results for direct reads (checking if the bit state inversions have been applied immediately after the inversion request was sent) can be found in Table~\ref{table:functions_experiment}. High-level results for delayed reads (checking if the bit state inversions have been retained after the 5 second Read 2 Time has elapsed) can be found in Table~\ref{table:functions_experiment_delay}. We can see that for each of the three configuration options, the quantity of vulnerable bytes differs. We can also see there is little difference between direct and delayed reads, aside from the MODBUSPN function. This means, in a holistic sense, that we can manipulate the behavior of FBs via their dedicated memory (fVBs). While our experiment was limited to inverting each bit within a byte, it highlights vulnerable areas of memory that an attack could target in a more elaborate way.

Overall, based on the FBs selected as part of our experiment, we can see that the use of direct inputs and gVBs does reduce the number of vulnerable bytes by around 6\%, though this is not a significant difference, as 90\% of bytes remain vulnerable. However, what is important to note is that for specific functions (GENERATE\_PULSE, WRITE\_DATA, and REMOTE\_CONN) a significant drop in vulnerable bytes was observed when not using default values. The PORT\_CONFIG\_DB function also saw a notable reduction from 97.7\% vulnerable bytes with default values, to 73.3\% vulnerable bytes with direct inputs, and 45.3\% vulnerable bytes with gVB inputs. Excluding the use of default values can therefore help in reducing vulnerabilities in control logic; this is a PLC programming practice that should be applied as part of baseline defensive measures.

\begin{table*}
\centering
\begin{tabular}{ c | c c c c c} 
\hline
\textbf{Function} & \textbf{Bytes} & \textbf{Variables} & \textbf{Defaults} & \textbf{Direct Input} & \textbf{gVBs} \\
\hline
\textbf{MODBUSPN} & 1544 & 82 & 1484 & 1444 & 1470 \\ 
\textbf{AS\_MAIL} & 990 & 42 & 977 & 889 & 889 \\
\textbf{CTU} & 9 & 10 & 6 & 4 & 4 \\ 
\textbf{CTD} & 9 & 10 & 6 & 4 & 4 \\ 
\textbf{TP} & 23 & 11 & 4 & 0 & 0 \\
\textbf{WRREC} & 26 & 12 & 20 & 4 & 4\\
\textbf{TSEND} & 22 & 14 & 20 & 8 & 8 \\
\textbf{PUT} & 598 & 106 & 594 & 572 & 572 \\
\textbf{TCON} & 20 & 13 & 18 & 6 & 6 \\ 
\textbf{PORT\_CONFIG} & 86 & 39 & 84 & 63 & 39 \\ 
\hline
\textbf{TOTAL} & 3327 & 339 & 3213 & 2994 & 2990\\ 
\hline
\end{tabular}
\vspace{5pt}
\caption{Overview of Tested Library Functions - Direct Read}
\label{table:functions_experiment}
\end{table*}

\begin{table*}
\centering
\begin{tabular}{ c | c c c c c} 
\hline
\textbf{Function} & \textbf{Bytes} & \textbf{Variables} & \textbf{Defaults} & \textbf{Direct Inputs} & \textbf{gVBs} \\
\hline
\textbf{MODBUSPN} & 1544 & 82 & 1482 & 1439 & 1465 \\ 
\textbf{AS\_MAIL} & 990 & 42 & 977 & 889 & 889\\
\textbf{CTU} & 9 & 10 & 6 & 4 & 4\\
\textbf{CTD} & 9 & 10 & 6 & 4 & 4\\
\textbf{TP} & 23 & 11 & 4 & 0 & 0\\
\textbf{WRREC} & 26 & 12 & 20 & 4 & 4\\
\textbf{TSEND} & 22 & 14 & 20 & 8 & 8\\
\textbf{PUT} & 598 & 106 & 594 & 572 & 572\\
\textbf{TCON} & 20 & 13 & 18 & 6 & 6\\ 
\textbf{PORT\_CONFIG} & 86 & 39 & 84 & 63 & 39\\ 
\hline
\textbf{TOTAL} & 3327 & 339 & 3211 & 2989 & 2991\\ 
\hline
\end{tabular}
\vspace{5pt}
\caption{Overview of Tested Library Functions - 5 Second Delayed Read}
\label{table:functions_experiment_delay}
\end{table*}

Taking the COUNT\_UP FB as an example, when we examine the scan results in more depth (see Table~\ref{table:COUNTUP}), we can evaluate what variables the vulnerable bits and bytes are mapped to, and what impact this could have on its operations should they be targeted by an attacker. Within COUNT\_UP several critical variables can be manipulated across the three configuration types, including CU (counter input), which is used as a trigger for the counter to count up, and CV (count value), which is the current count value. Only the PV variable changes from vulnerable to non-vulnerable when a direct input or gVB input is used compared to default values. An adversary could use these vulnerable variables to stop the control logic from counting, or modify the overall count value with a single request (S7Comms Write). An attack of this nature could prevent the control logic from performing additional critical operations, such as notifying an engineer with an alarm when a given count limit has been reached. Further to this, even though Q (counter output) is not vulnerable, an adversary could manipulate this boolean variable by manipulating the input bytes which are vulnerable. Fundamentally, once the scanner has returned its results, an analysis similar to this must be undertaken by the user to evaluate the list of vulnerable bits and bytes, and identify the impact of their manipulation. This is a critical task, where results are fed into risk assessment processes, with a goal of bridging the vulnerability-impact gap.

\begin{table}
\begin{tabular}{ l | c | c | l} 
\hline
\textbf{Variable} & \textbf{Vuln.} & \textbf{Type} & \textbf{Description}\\
\hline
\textbf{CU} & \checkmark & Bool & Counter input\\ 
\textbf{R} & \checkmark & Bool & Reset\\ 
\textbf{PV} & (\checkmark) & Int & Max count before Q is triggered\\
\textbf{Q} &  & Bool & Indicates if CV is greater than PV\\
\textbf{CV} & \checkmark & Int & Count value\\
\end{tabular}
\vspace{5pt}
\caption{COUNT\_UP}
\label{table:COUNTUP}
\end{table}

During our experimentation we established that pointer-based inputs are less susceptible to attack (e.g., MODBUSPN where variables storing pointer addresses could not be inverted). However, this does not mean the variable is not indirectly vulnerable, as it could point to a vulnerable area of memory. We have undertaken a proof-of-concept attack to examine this possibility further (see Figure~\ref{fig:pointer}), by first reading the pointer address (a gVB address) stored in a fVB linked to the REG\_KEY variable (this stores a registration key for use by the MODBUSPN function). Once the gVB address has been obtained, we are able to target it and successfully invert the state of its bits. This bit inversion was then observed by the fVB.


Our scanner was able to identify these previously unknown vulnerabilities within the Siemens TIA libraries, which could enable an attacker to manipulate fVB data and therefore impact FB behavior. This could create wide-spread impact to physical operational processes, communication and alerting functions, and more.

We also ran Nessus and OpenVAS on the same PLC used for this evaluation. These tools identified vulnerabilities such as CVE-1999-0517 (related to SNMP); however, they did not offer insight into any control logic derived vulnerabilities, or the targeted manipulation of control logic behaviour. Additionally, the use of Nessus did cause our PLC to crash during one scan. The results provided by both tools offered basic high-level generic vulnerability information, thus failing to bridge the vulnerability-impact gap in the same way as our approach.

\subsection{Example Attack Scenario}
Based on our proof of concept, performed on Siemens library functions, there are several attack vectors based on our identified control logic vulnerability that could be identified. As we are sending unauthorised command messages to the PLC in order to change the state of the memory and impact control logic, we can change the behaviour of the PLC. What can be changed depends on the function and how the variables are written to it. Looking at four library functions (COUNT\_UP, AS\_MAIL, GENERATE\_PULSE and TCON), we can explore one of the potential attack scenarios.

If we have a program that utilises the COUNT\_UP function to count (CV) how many times a valve has opened (CU) within a specific period which is reset when the GENERATE\_PULSE function sends a pulse to the COUNT\_UP function (R). When the counter reaches 10 (PV), a pulse (Q) is sent to the AS\_MAIL and TCON functions. The AS\_MAIL function, within this example, is configured to send an email to an engineer and the operations centre, and the TCON function writes data to a web server over TCP. An attacker would only need to interfere with the operation of the COUNT\_UP function to interfere with the operation of the other two functions and by proxy the operation of the PLC and the associated alarm reporting processes. By continuously writing a FALSE value to the CU variable, the function might never trigger the counter input; therefore, the count value (CV) will not increase when it should. Given that the CV variable is also vulnerable, the attacker could continuously write 0 to this value so it would never reach the max count (PV) value that is required for the function to send a pulse to AS\_MAIL and TCON. An attacker could also repeatedly write a TRUE value to the reset (R) variable, which would reset the count value to 0. Finally, the attacker could also interfere with the max count (PV) value if this value was left as default (0), which we deem to be unlikely as an engineer would set this value to enable the function to perform. An attack like this could allow other attacks to be executed without alerting the OT engineer, as reporting processes would be impacted and could eventually result in the activation of safety systems which could cause the plant to shut down.

However, an adversary with access to the PLC could adversely impact the function in multiple ways by exploiting this vulnerability. All three other functions have vulnerable bytes, which can also be used to affect the operation of this example program. For example, the BUSY variable for the TCON and AS MAIL function is vulnerable; this variable is used to identify if the function is busy, and if so, a new job cannot be started. If an attacker overwrites the BUSY variable with a TRUE value, the PLC will halt the function's operation. Some of the vulnerable bytes can become non-vulnerable if direct inputs or gVB is used to provide values. Therefore, good programming practice might mitigate some of the attack vectors.

\section{Lessons Learned}
\label{discussion}
As observed from the results of our experiment, we can see that identifying vulnerable areas of PLC memory offers a new perspective on vulnerability scanning that is not currently provided by traditional scanning techniques. If PLC memory can be manipulated via the use of unauthorised command messages, malicious actors are able to change the behaviour of control logic, leading to undesirable operational impact. From scanning 10 distinct library fVBs, we concluded that even vendor provided control logic can be vulnerable. This therefore raises further concerns; more specifically, if vendor provided library functions cannot be considered secure against memory manipulation attacks, what level of vulnerabilities could be exploited within code written by PLC engineers, who are less likely to follow secure programming practices? Not only would this code rely on vulnerable vendor written libraries, but also on self-written sections of code.

To mitigate a number of the identified vulnerable bytes discussed in Section~\ref{results}, we have identified that input parameters should always be set (not left as default), as this results in fewer vulnerable bytes. When left unset (default) these values are only set when the PLC initializes, which makes it easy for the values to be overwritten with persistence. The use of concepts such as 'stack canaries' and other similar alerts could also be adopted to monitor predictable data structures indicating the presence of unauthorized command messages targeting vulnerable memory areas. However, at a fundamental level, the control logic itself needs to be reviewed to better understand why certain areas of memory and their associated variables are vulnerable to begin with, and whether modifications can be made to mitigate the risk within the baseline code itself (as described in the example from Sections~\ref{scanning_cl} and~\ref{scanning_mem}).

\subsection{The Impact of PLC-VBS on cyber risk assessment}
What we aim to achieve with the development of the PLC-VBS scanner is to improve the understanding of vulnerable PLC control logic, something we believe would be a valuable addition to both cyber risk assessment and evaluating the efficacy of cyber security controls. Traditionally, such endeavors of cyber risk assessment in an ICS environment would rely on vulnerability scanning tools designed for the distinctly different environments of enterprise IT, rendering them less effective, or even ineffective \citep{nessus,openvas} (as we found during our experimentation). Such non-specific vulnerability scanning tools do not investigate a PLC's memory, often stopping at discerning the ports open on a device, and perhaps the services running on those ports. Conversely, using the tool we have presented would allow cyber risk assessment practitioners to delve deeper into the vulnerabilities within the PLC control logic itself, providing granular contextual data and the ability to link back to physical operational processes (e.g., the identification of vulnerabilities in control logic responsible for controlling a valve to maintain a water tank level), and therefore enrich the output of the cyber risk assessment for decision-making recipients. 

\subsection{Cross Vendor Generalization}
Although we have focused on Siemens PLCs and use of the S7 protocol within our proof of concept, unauthorised command message attacks are equally applicable to other vendors, such as Allen Bradley~\citep{green2021}, and protocols, such as Modbus. As HMIs operate using memory access (see Figure~\ref{fig:process}), we can shift the focus of this scanner from vendors to protocols. In its current form our scanner can be used against any device that leverages the S7 protocol; however, it could be expanded on by implementing support for additional protocols such as Modbus or OPC Unified Architecture through existing corresponding libraries (pymodbus or opcua-asyncio respectively~\citep{pymodbus, opcua-asyncio}).

\section{Conclusion and Future Work}
\label{conclusion}
In this paper we have introduced and evaluated PLC-VBS. We have shown that our scanner is able to identify vulnerable bits and bytes within PLC memory, through which adversaries can interact with the device and its associated underlying operational processes. During the evaluation we tested 10 vendor-provided library functions, and identified that they are vulnerable to manipulation through the use of unauthorised command message attacks. This proves that anyone able to access the PLC over the network can change values within memory reserved for commonly used functions. This could result in a disastrous impact and allow adversaries to interact with the PLC while hiding alerts from system operators. For example, if operators are to be notified when a counter reaches a certain value, an adversary could keep resetting this to a lower value, and therefore prevent alerts from being generated, while also impacting other areas of control logic that may require this value to operate (e.g., to start a pump when the count reaches a given threshold).

We believe PLC-VBS can be used to enhance risk assessment activities compared to conventional IT-system focused scanners (e.g., Nessus and OpenVAS), and should inform PLC programming practices to provide device-level mitigation. For example, it is important to take into account that keeping default values as FB inputs leads to more vulnerable FBs, when compared to the use of direct or gVB inputs. By identifying and addressing control logic vulnerabilities we are able to provide a more complete and holistic level of defensive measures. However, other appropriate mitigation techniques such as network-based detection and honeypots~\citep{maesschalck2022} can also be provided to deliver a defence-in-depth strategy. For example, honeypots could be used to identify if adversaries are actively exploiting the identified vulnerabilities when configured with appropriate levels of interactivity~\citep{maesschalck2021}. To begin mitigating the vulnerabilities discussed in this paper we would encourage PLC engineers to avoid reliance on default input values and public libraries. This could form part of a devsecops approach, where security is an inherent part of the development and operational process. In addition, we advise protection of VBs from direct manipulation over the network where this level of connectivity is not required for operational purposes (e.g., HMI access).

\begin{figure}
	\begin{center}
    	\includegraphics[width=1\columnwidth]{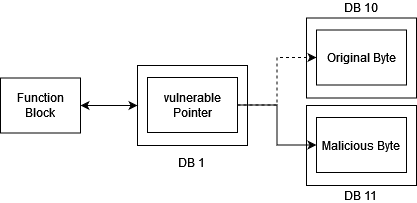}
	\end{center}
    \caption{Pointer Re-direction Example}
    \label{fig:pointer1}
\end{figure}

While we have shown that there is a multitude of vulnerable bytes within the tested fVBs, potential misuse of these is not necessarily avoided when a byte is not vulnerable. We can also identify pointer addresses within functions that are potentially vulnerable to being overwritten, which would then point to an address of the adversary's choosing, i.e. performing a pointer re-direction. This can be seen in Figure~\ref{fig:pointer1} where we show a function with a pointer in DB1 that used to point to an address in DB 10 but is now redirected to point to an address in DB 11.

\begin{figure}
	\begin{center}
    	\includegraphics[width=1\columnwidth]{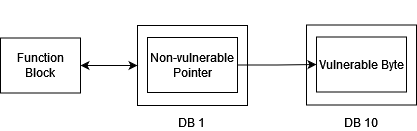}
	\end{center}
    \caption{Indirect Vulnerable Pointer Example}
    \label{fig:pointer}
\end{figure}

Further to this, if the pointer itself cannot be overwritten, the address it is pointing to might be vulnerable. Hence, the pointer can be viewed as vulnerable by proxy. Furthermore, if we can identify multiple pointer addresses across VBs pointing towards the same gVB, we can establish the "primary gVB"/gVB of high importance across the control logic base. We can then investigate this VB (e.g., DB 10 in Figure~\ref{fig:pointer}) and monitor changes across multiple other VBs (e.g. DB 1, DB 2, DB 3, and DB 4) potentially interfering with previously non-vulnerable addresses. However, it is important to note that when we refer to pointers, we only have the pointer start address, and would need to establish the length of data stored at that address. An example of this can be seen in Figure~\ref{fig:pointer} where we depict a function that stores values in DB 1 and has a pointer that points to a value stored in DB 10.

Additionally, we have not investigated whether bits and/or bytes (non-pointers) are vulnerable by association to other bits and bytes. How this future work fits into PLC-VBS is indicated in Figure~\ref{fig:tool}; this will focus on addresses of high importance, such as "plant shutdown", which might be protected against overwriting directly. However, flipping other bits might affect this address. For example, bit C might not be vulnerable but bits A and B are; while we cannot flip bit C, we could investigate if bit C changes state after flipping bit A and B. This would mean bit C is indirectly vulnerable. If so we could then focus our efforts on simultaneously manipulating multiple variables within a VB to change the state of operator-specified critical variables.

Other future work will cover additional network protocols to target different devices, and the previously noted extensions of work to explore indirect address manipulation vulnerabilities (including the role of pointers).



\section*{Acknowledgements}
The authors gratefully acknowledge the support of the Next
Generation Converged Digital Infrastructure (NG-CDI) Prosperity
Partnership funded by UK’s EPSRC and British Telecom plc (award
number EP/R004935/1)

\printcredits

\bibliographystyle{cas-model2-names}

\bibliography{refs}

\begin{thebibliography}{51}
\expandafter\ifx\csname natexlab\endcsname\relax\def\natexlab#1{#1}\fi
\providecommand{\url}[1]{\texttt{#1}}
\providecommand{\href}[2]{#2}
\providecommand{\path}[1]{#1}
\providecommand{\DOIprefix}{doi:}
\providecommand{\ArXivprefix}{arXiv:}
\providecommand{\URLprefix}{URL: }
\providecommand{\Pubmedprefix}{pmid:}
\providecommand{\doi}[1]{\href{http://dx.doi.org/#1}{\path{#1}}}
\providecommand{\Pubmed}[1]{\href{pmid:#1}{\path{#1}}}
\providecommand{\bibinfo}[2]{#2}
\ifx\xfnm\relax \def\xfnm[#1]{\unskip,\space#1}\fi
\bibitem[{Abbasi et~al.(2016)Abbasi, Hashemi, Zambon and
  Etalle}]{abbasi2016ghost}
\bibinfo{author}{Abbasi, A.}, \bibinfo{author}{Hashemi, M.},
  \bibinfo{author}{Zambon, E.}, \bibinfo{author}{Etalle, S.},
  \bibinfo{year}{2016}.
\newblock \bibinfo{title}{Stealth low-level manipulation of programmable logic
  controllers i/o by pin control exploitation}, in:
  \bibinfo{booktitle}{International Conference on Critical Information
  Infrastructures Security}, \bibinfo{publisher}{Springer},
  \bibinfo{address}{New York}. pp. \bibinfo{pages}{1--12}.
\bibitem[{Ahmad et~al.(2019)Ahmad, Webb, Desouza and Boorman}]{ahmad2019}
\bibinfo{author}{Ahmad, A.}, \bibinfo{author}{Webb, J.},
  \bibinfo{author}{Desouza, K.C.}, \bibinfo{author}{Boorman, J.},
  \bibinfo{year}{2019}.
\newblock \bibinfo{title}{Strategically-motivated advanced persistent threat:
  Definition, process, tactics and a disinformation model of counterattack}.
\newblock \bibinfo{journal}{Computers \& Security} \bibinfo{volume}{86},
  \bibinfo{pages}{402--418}.
\bibitem[{Ahmed et~al.(2017)Ahmed, Obermeier, Sudhakaran and
  Roussev}]{Ahmed2017}
\bibinfo{author}{Ahmed, I.}, \bibinfo{author}{Obermeier, S.},
  \bibinfo{author}{Sudhakaran, S.}, \bibinfo{author}{Roussev, V.},
  \bibinfo{year}{2017}.
\newblock \bibinfo{title}{{Programmable Logic Controller Forensics}}.
\newblock \bibinfo{journal}{IEEE Security and Privacy} \bibinfo{volume}{15},
  \bibinfo{pages}{18--24}.
\bibitem[{{Allen Bradley}(2008)}]{AllenBradley2008}
\bibinfo{author}{{Allen Bradley}}, \bibinfo{year}{2008}.
\newblock \bibinfo{title}{{SLC 500 Instruction Set}}.
\newblock \bibinfo{type}{Technical Report}. Allan Bradley.
\newblock \URLprefix
  \url{https://literature.rockwellautomation.com/idc/groups/literature/documents/rm/1747-rm001{\_}-en-p.pdf}.
\bibitem[{Antrobus et~al.(2016)Antrobus, Frey, Green and Rashid}]{antrobus2016}
\bibinfo{author}{Antrobus, R.}, \bibinfo{author}{Frey, S.},
  \bibinfo{author}{Green, B.}, \bibinfo{author}{Rashid, A.},
  \bibinfo{year}{2016}.
\newblock \bibinfo{title}{Simaticscan: Towards a specialised vulnerability
  scanner for industrial control systems}, in: \bibinfo{booktitle}{4th
  International Symposium for ICS \& SCADA Cyber Security Research 2016 4},
  \bibinfo{publisher}{BCS Learning \& Development Ltd.},
  \bibinfo{address}{Swindon}. pp. \bibinfo{pages}{11--18}.
\bibitem[{Antrobus et~al.(2019)Antrobus, Green, Frey and Rashid}]{antrobus2019}
\bibinfo{author}{Antrobus, R.}, \bibinfo{author}{Green, B.},
  \bibinfo{author}{Frey, S.}, \bibinfo{author}{Rashid, A.},
  \bibinfo{year}{2019}.
\newblock \bibinfo{title}{The forgotten i in iiot: a vulnerability scanner for
  industrial internet of things}, in: \bibinfo{booktitle}{IET Conference
  Proceedings}, \bibinfo{publisher}{IET}, \bibinfo{address}{London}.
\bibitem[{Biham et~al.(2019)Biham, Bitan, Carmel, Dankner, Malin and
  Woo}]{Biham2019}
\bibinfo{author}{Biham, E.}, \bibinfo{author}{Bitan, S.},
  \bibinfo{author}{Carmel, A.}, \bibinfo{author}{Dankner, A.},
  \bibinfo{author}{Malin, U.}, \bibinfo{author}{Woo, A.}, \bibinfo{year}{2019}.
\newblock \bibinfo{title}{{Rogue7: Rogue Engineering-Station Attacks on
  S7Simatic PLCs}}.
\newblock \bibinfo{type}{Technical Report}. Black Hat USA.
\bibitem[{Bristow(2008)}]{Bristow2008}
\bibinfo{author}{Bristow, M.}, \bibinfo{year}{2008}.
\newblock \bibinfo{title}{{ModScan - A SCADA MODBUS Network Scanner}}.
\newblock
  \bibinfo{howpublished}{\url{https://defcon.org/images/defcon-16/dc16-presentations/defcon-16-bristow.pdf}}.
\bibitem[{{British Standards Institute}(2013)}]{BritishStandardsInstitute2013}
\bibinfo{author}{{British Standards Institute}}, \bibinfo{year}{2013}.
\newblock \bibinfo{title}{{61132-3:2013 - Programmable Controllers - Part 3:
  Programming Languages}}.
\newblock \bibinfo{type}{Technical Report}. BSI.
\newblock \URLprefix
  \url{https://bsol.bsigroup.com/Search/Search?searchKey=61131-3{\&}OriginPage=Header+Search+Box{\&}autoSuggestion=false}.
\bibitem[{{Centre for the Protection of National
  Infrastructure}(2021)}]{CPNI2021}
\bibinfo{author}{{Centre for the Protection of National Infrastructure}},
  \bibinfo{year}{2021}.
\newblock \bibinfo{title}{{Critical National Infrastructure}}.
\newblock
  \bibinfo{howpublished}{\url{https://www.cpni.gov.uk/critical-national-infrastructure-0}}.
\newblock \bibinfo{note}{Last Accessed: 2022-06-16}.
\bibitem[{CPNI(2020)}]{CPNI2020}
\bibinfo{author}{CPNI}, \bibinfo{year}{2020}.
\newblock \bibinfo{title}{{Critical National Infrastructure}}.
\newblock
  \bibinfo{howpublished}{\url{https://www.cpni.gov.uk/critical-national-infrastructure-0}}.
\newblock \URLprefix
  \url{https://www.cpni.gov.uk/critical-national-infrastructure-0}.
\bibitem[{{Cybersecurity \& Infrastructure Security Agency}()}]{CISA-HMI}
\bibinfo{author}{{Cybersecurity \& Infrastructure Security Agency}}, .
\newblock \bibinfo{title}{{Control System HMI Computers}}.
\newblock
  \bibinfo{howpublished}{\url{https://www.cisa.gov/uscert/ics/Control_System_HMI_Computers-Definition.html}}.
\newblock \bibinfo{note}{Last Accessed: 2022-06-24}.
\bibitem[{Derbyshire et~al.(2021)Derbyshire, Green and Hutchison}]{cost}
\bibinfo{author}{Derbyshire, R.}, \bibinfo{author}{Green, B.},
  \bibinfo{author}{Hutchison, D.}, \bibinfo{year}{2021}.
\newblock \bibinfo{title}{{“Talking a different Language”: Anticipating
  adversary attack cost for cyber risk assessment}}.
\newblock \bibinfo{journal}{Computers \& Security} \bibinfo{volume}{103},
  \bibinfo{pages}{102163}.
\bibitem[{Derbyshire et~al.(2018)Derbyshire, Green, Prince, Mauthe and
  Hutchison}]{Derbyshire2018}
\bibinfo{author}{Derbyshire, R.}, \bibinfo{author}{Green, B.},
  \bibinfo{author}{Prince, D.}, \bibinfo{author}{Mauthe, A.},
  \bibinfo{author}{Hutchison, D.}, \bibinfo{year}{2018}.
\newblock \bibinfo{title}{{An Analysis of Cyber Security Attack Taxonomies}},
  in: \bibinfo{booktitle}{Security and Privacy Workshops (EuroS{\&}PW), 2018
  IEEE European Symposium on}, \bibinfo{publisher}{IEEE}, \bibinfo{address}{New
  Jersey}. pp. \bibinfo{pages}{153--161}.
\bibitem[{Didier et~al.(2011)Didier, Macias, Harstad, Antholine, Johnston,
  Piyevsky, Zaniewski, Zuponcic, Schillace and Wilcox}]{Didier2011}
\bibinfo{author}{Didier, P.}, \bibinfo{author}{Macias, F.},
  \bibinfo{author}{Harstad, J.}, \bibinfo{author}{Antholine, R.},
  \bibinfo{author}{Johnston, S.A.}, \bibinfo{author}{Piyevsky, S.},
  \bibinfo{author}{Zaniewski, D.}, \bibinfo{author}{Zuponcic, S.},
  \bibinfo{author}{Schillace, M.}, \bibinfo{author}{Wilcox, G.},
  \bibinfo{year}{2011}.
\newblock \bibinfo{title}{{Converged Plantwide Ethernet (CPwE) Design and
  Implementation Guide}}.
\newblock \bibinfo{journal}{Rockwell Automation} \bibinfo{volume}{9},
  \bibinfo{pages}{564}.
\bibitem[{Drias et~al.(2015)Drias, Serhrouchni and Vogel}]{drias2015taxonomy}
\bibinfo{author}{Drias, Z.}, \bibinfo{author}{Serhrouchni, A.},
  \bibinfo{author}{Vogel, O.}, \bibinfo{year}{2015}.
\newblock \bibinfo{title}{Taxonomy of attacks on industrial control protocols},
  in: \bibinfo{booktitle}{2015 International Conference on Protocol Engineering
  (ICPE) and International Conference on New Technologies of Distributed
  Systems (NTDS)}, \bibinfo{publisher}{IEEE}, \bibinfo{address}{New Jersey}.
  pp. \bibinfo{pages}{1--6}.
\bibitem[{Eckhart et~al.(2019)Eckhart, Ekelhart, L{\"u}der, Biffl and
  Weippl}]{eckhart2019security}
\bibinfo{author}{Eckhart, M.}, \bibinfo{author}{Ekelhart, A.},
  \bibinfo{author}{L{\"u}der, A.}, \bibinfo{author}{Biffl, S.},
  \bibinfo{author}{Weippl, E.}, \bibinfo{year}{2019}.
\newblock \bibinfo{title}{Security development lifecycle for cyber-physical
  production systems}, in: \bibinfo{booktitle}{IECON 2019-45th Annual
  Conference of the IEEE Industrial Electronics Society},
  \bibinfo{publisher}{IEEE}, \bibinfo{address}{New Jersey}. pp.
  \bibinfo{pages}{3004--3011}.
\bibitem[{Erickson(1996)}]{erickson1996}
\bibinfo{author}{Erickson, K.T.}, \bibinfo{year}{1996}.
\newblock \bibinfo{title}{{Programmable logic Controllers}}.
\newblock \bibinfo{journal}{IEEE Potentials} \bibinfo{volume}{15},
  \bibinfo{pages}{14--17}.
\bibitem[{{European Commission}(2019)}]{EuropeanCommission2019}
\bibinfo{author}{{European Commission}}, \bibinfo{year}{2019}.
\newblock \bibinfo{title}{{The Directive on security of network and information
  systems (NIS Directive)}}.
\newblock
  \bibinfo{howpublished}{\url{https://ec.europa.eu/digital-single-market/en/network-and-information-security-nis-directive}}.
\newblock \URLprefix
  \url{https://ec.europa.eu/digital-single-market/en/network-and-information-security-nis-directive}.
\bibitem[{Feld(2004)}]{feld2004profinet}
\bibinfo{author}{Feld, J.}, \bibinfo{year}{2004}.
\newblock \bibinfo{title}{Profinet-scalable factory communication for all
  applications}, in: \bibinfo{booktitle}{IEEE International Workshop on Factory
  Communication Systems, 2004. Proceedings.}, \bibinfo{publisher}{IEEE},
  \bibinfo{address}{New Jersey}. pp. \bibinfo{pages}{33--38}.
\bibitem[{Fluchs(2020)}]{Fluchs2020}
\bibinfo{author}{Fluchs, S.}, \bibinfo{year}{2020}.
\newblock \bibinfo{title}{{The Top 20 Secure PLC Coding Practices Project}}.
\newblock \bibinfo{howpublished}{\url{https://bit.ly/34DqoHi}}.
\bibitem[{Green et~al.(2020)Green, Derbyshire, Knowles, Boorman, Ciholas,
  Prince and Hutchison}]{green2020}
\bibinfo{author}{Green, B.}, \bibinfo{author}{Derbyshire, R.},
  \bibinfo{author}{Knowles, W.}, \bibinfo{author}{Boorman, J.},
  \bibinfo{author}{Ciholas, P.}, \bibinfo{author}{Prince, D.},
  \bibinfo{author}{Hutchison, D.}, \bibinfo{year}{2020}.
\newblock \bibinfo{title}{$\{$ICS$\}$ testbed tetris: Practical building blocks
  towards a cyber security resource}, in: \bibinfo{booktitle}{13th
  $\{$USENIX$\}$ Workshop on Cyber Security Experimentation and Test
  ($\{$CSET$\}$ 20)}, \bibinfo{publisher}{{USENIX}},
  \bibinfo{address}{Berkeley, CA}.
\bibitem[{Green et~al.(2021)Green, Derbyshire, Krotofil, Knowles, Prince and
  Suri}]{green2021}
\bibinfo{author}{Green, B.}, \bibinfo{author}{Derbyshire, R.},
  \bibinfo{author}{Krotofil, M.}, \bibinfo{author}{Knowles, W.},
  \bibinfo{author}{Prince, D.}, \bibinfo{author}{Suri, N.},
  \bibinfo{year}{2021}.
\newblock \bibinfo{title}{{PCaaD}: Towards automated determination and
  exploitation of industrial systems}.
\newblock \bibinfo{journal}{Computers \& Security} \bibinfo{volume}{110},
  \bibinfo{pages}{102424}.
\bibitem[{Green et~al.(2017a)Green, Krotofil and Abbasi}]{Green2017a}
\bibinfo{author}{Green, B.}, \bibinfo{author}{Krotofil, M.},
  \bibinfo{author}{Abbasi, A.}, \bibinfo{year}{2017}a.
\newblock \bibinfo{title}{{On the Significance of Process Comprehension for
  Conducting Targeted ICS Attacks}}, in: \bibinfo{booktitle}{Proceedings of the
  3nd ACM Workshop on Cyber-Physical Systems Security and Privacy},
  \bibinfo{publisher}{ACM}, \bibinfo{address}{New York}. pp.
  \bibinfo{pages}{57--67}.
\bibitem[{Green et~al.(2017b)Green, Lee, Antrobus, Roedig, Hutchison and
  Rashid}]{Green2017}
\bibinfo{author}{Green, B.}, \bibinfo{author}{Lee, A.},
  \bibinfo{author}{Antrobus, R.}, \bibinfo{author}{Roedig, U.},
  \bibinfo{author}{Hutchison, D.}, \bibinfo{author}{Rashid, A.},
  \bibinfo{year}{2017}b.
\newblock \bibinfo{title}{{Pains, Gains and PLCs: Ten Lessons from Building an
  Industrial Control Systems Testbed for Security Research}}, in:
  \bibinfo{booktitle}{10th USENIX Workshop on Cyber Security Experimentation
  and Test (CSET 17)}, \bibinfo{publisher}{USENIX Association}.
\bibitem[{Greenbone(2022)}]{openvas}
\bibinfo{author}{Greenbone}, \bibinfo{year}{2022}.
\newblock \bibinfo{title}{Openvas - open vulnerability assessment scanner}.
\newblock \bibinfo{howpublished}{\url{https://www.openvas.org/}}.
\bibitem[{Kottler et~al.(2017)Kottler, Khayamy, Hasan and
  Elkeelany}]{kottler2017formal}
\bibinfo{author}{Kottler, S.}, \bibinfo{author}{Khayamy, M.},
  \bibinfo{author}{Hasan, S.R.}, \bibinfo{author}{Elkeelany, O.},
  \bibinfo{year}{2017}.
\newblock \bibinfo{title}{Formal verification of ladder logic programs using
  nusmv}, in: \bibinfo{booktitle}{SoutheastCon 2017},
  \bibinfo{publisher}{IEEE}, \bibinfo{address}{New Jersey}. pp.
  \bibinfo{pages}{1--5}.
\bibitem[{Library(2021)}]{opcua-asyncio}
\bibinfo{author}{Library, F.O.U.}, \bibinfo{year}{2021}.
\newblock \bibinfo{title}{Opc ua / iec 62541 client and server for python}.
\newblock
  \bibinfo{howpublished}{\url{https://github.com/FreeOpcUa/opcua-asyncio}}.
\bibitem[{Ljungkrantz and Akesson(2007)}]{Ljungkrantz2007}
\bibinfo{author}{Ljungkrantz, O.}, \bibinfo{author}{Akesson, K.},
  \bibinfo{year}{2007}.
\newblock \bibinfo{title}{{A study of industrial logic control programming
  using library components}}, in: \bibinfo{booktitle}{2007 IEEE International
  Conference on Automation Science and Engineering}, \bibinfo{publisher}{IEEE},
  \bibinfo{address}{New Jersey}. pp. \bibinfo{pages}{117--122}.
\bibitem[{Maesschalck et~al.(2022)Maesschalck, Giotsas, Green and
  Race}]{maesschalck2022}
\bibinfo{author}{Maesschalck, S.}, \bibinfo{author}{Giotsas, V.},
  \bibinfo{author}{Green, B.}, \bibinfo{author}{Race, N.},
  \bibinfo{year}{2022}.
\newblock \bibinfo{title}{Don’t get stung, cover your {ICS} in honey: How do
  honeypots fit within industrial control system security}.
\newblock \bibinfo{journal}{Computers \& Security} \bibinfo{volume}{114},
  \bibinfo{pages}{102598}.
\bibitem[{Maesschalck et~al.(2021)Maesschalck, Giotsas and
  Race}]{maesschalck2021}
\bibinfo{author}{Maesschalck, S.}, \bibinfo{author}{Giotsas, V.},
  \bibinfo{author}{Race, N.}, \bibinfo{year}{2021}.
\newblock \bibinfo{title}{World wide {ICS} honeypots: A study into the
  deployment of conpot honeypots}, in: \bibinfo{booktitle}{Seventh Annual
  Industrial Control System Security (ICSS) Workshop},
  \bibinfo{publisher}{ACSAC}, \bibinfo{address}{Maryland}.
\bibitem[{McLaughlin et~al.(2016)McLaughlin, Konstantinou, Wang, Davi, Sadeghi,
  Maniatakos and Karri}]{McLaughlin2016}
\bibinfo{author}{McLaughlin, S.}, \bibinfo{author}{Konstantinou, C.},
  \bibinfo{author}{Wang, X.}, \bibinfo{author}{Davi, L.},
  \bibinfo{author}{Sadeghi, A.R.}, \bibinfo{author}{Maniatakos, M.},
  \bibinfo{author}{Karri, R.}, \bibinfo{year}{2016}.
\newblock \bibinfo{title}{{The Cybersecurity Landscape in Industrial Control
  Systems}}.
\newblock \bibinfo{journal}{Proceedings of the IEEE} \bibinfo{volume}{104},
  \bibinfo{pages}{1039--1057}.
\bibitem[{McMahon et~al.(2018)McMahon, Patton, Samtani and Chen}]{mcmahon2018}
\bibinfo{author}{McMahon, E.}, \bibinfo{author}{Patton, M.},
  \bibinfo{author}{Samtani, S.}, \bibinfo{author}{Chen, H.},
  \bibinfo{year}{2018}.
\newblock \bibinfo{title}{Benchmarking vulnerability assessment tools for
  enhanced cyber-physical system (cps) resiliency}, in:
  \bibinfo{booktitle}{2018 IEEE International Conference on Intelligence and
  Security Informatics (ISI)}, \bibinfo{publisher}{IEEE}, \bibinfo{address}{New
  Jersey}. pp. \bibinfo{pages}{100--105}.
\bibitem[{Miller et~al.(2021)Miller, Staves, Maesschalck, Sturdee and
  Green}]{miller2021}
\bibinfo{author}{Miller, T.}, \bibinfo{author}{Staves, A.},
  \bibinfo{author}{Maesschalck, S.}, \bibinfo{author}{Sturdee, M.},
  \bibinfo{author}{Green, B.}, \bibinfo{year}{2021}.
\newblock \bibinfo{title}{Looking back to look forward: Lessons learnt from
  cyber-attacks on industrial control systems}.
\newblock \bibinfo{journal}{International Journal of Critical Infrastructure
  Protection} \bibinfo{volume}{35}, \bibinfo{pages}{100464}.
\bibitem[{MITRE(2021)}]{MITRE2021}
\bibinfo{author}{MITRE}, \bibinfo{year}{2021}.
\newblock \bibinfo{title}{{Unauthorized Command Message}}.
\newblock
  \bibinfo{howpublished}{\url{https://collaborate.mitre.org/attackics/index.php/Technique/T0855}}.
\bibitem[{Molenaar and Preeker(2013)}]{Molenaar2013}
\bibinfo{author}{Molenaar, G.}, \bibinfo{author}{Preeker, S.},
  \bibinfo{year}{2013}.
\newblock \bibinfo{title}{{Welcome to python-snap7's documentation!}}
\newblock
  \bibinfo{howpublished}{\url{https://python-snap7.readthedocs.io/en/latest/}}.
\newblock \URLprefix \url{https://python-snap7.readthedocs.io/en/latest/}.
\bibitem[{Nochvay(2019)}]{Nochvay2019}
\bibinfo{author}{Nochvay, A.}, \bibinfo{year}{2019}.
\newblock \bibinfo{title}{{CODESYS Runtime, a PLC control framework}}.
\newblock \bibinfo{type}{Technical Report}. Kaspersky.
\bibitem[{PLCDev(2020)}]{PLCDev2020}
\bibinfo{author}{PLCDev}, \bibinfo{year}{2020}.
\newblock \bibinfo{title}{{Symbol Table Allowed Addresses and Data Types}}.
\newblock
  \bibinfo{howpublished}{\url{http://www.plcdev.com/symbol\_table\_allowed\_addresses\_and\_data\_types}}.
\bibitem[{Riptide(2021)}]{pymodbus}
\bibinfo{author}{Riptide}, \bibinfo{year}{2021}.
\newblock \bibinfo{title}{Pymodbus - a python modbus stack}.
\newblock \bibinfo{howpublished}{\url{https://github.com/riptideio/pymodbus}}.
\bibitem[{Robles-Durazno et~al.(2019)Robles-Durazno, Moradpoor, McWhinnie,
  Russell and Maneru-Marin}]{Robles-Durazno2019}
\bibinfo{author}{Robles-Durazno, A.}, \bibinfo{author}{Moradpoor, N.},
  \bibinfo{author}{McWhinnie, J.}, \bibinfo{author}{Russell, G.},
  \bibinfo{author}{Maneru-Marin, I.}, \bibinfo{year}{2019}.
\newblock \bibinfo{title}{{PLC memory attack detection and response in a clean
  water supply system}}.
\newblock \bibinfo{journal}{International Journal of Critical Infrastructure
  Protection} \bibinfo{volume}{26}, \bibinfo{pages}{100300}.
\bibitem[{Searle(2015)}]{PLCScan}
\bibinfo{author}{Searle, J.}, \bibinfo{year}{2015}.
\newblock \bibinfo{title}{Plcscan}.
\newblock \bibinfo{howpublished}{\url{https://github.com/meeas/plcscan}}.
\newblock \bibinfo{note}{Last accessed: 21 March 2022}.
\bibitem[{Shodan(2020)}]{Shodan2020}
\bibinfo{author}{Shodan}, \bibinfo{year}{2020}.
\newblock \bibinfo{title}{{Industrial Control Systems}}.
\newblock \bibinfo{howpublished}{\url{https://bit.ly/3Jh3kAG}}.
\bibitem[{Siemens(2019)}]{Siemens2019}
\bibinfo{author}{Siemens}, \bibinfo{year}{2019}.
\newblock \bibinfo{title}{{Library of general functions (LGF) for SIMATIC STEP
  7 (TIA Portal) and SIMATIC S7-1200 / S7-1500}}.
\newblock \bibinfo{howpublished}{\url{https://sie.ag/3aeTFKz}}.
\newblock \URLprefix \url{https://sie.ag/3aeTFKz}.
\bibitem[{Siemens(2020a)}]{Siemens2020a}
\bibinfo{author}{Siemens}, \bibinfo{year}{2020}a.
\newblock \bibinfo{title}{{SIMATIC S7-300 - Proven Multiple Times!}}
\newblock \bibinfo{howpublished}{\url{https://sie.ag/3ol428k}}.
\bibitem[{Siemens(2020b)}]{Siemenswincc}
\bibinfo{author}{Siemens}, \bibinfo{year}{2020}b.
\newblock \bibinfo{title}{{SIMATIC WinCC V7}}.
\newblock
  \bibinfo{howpublished}{\url{https://new.siemens.com/global/en/products/automation/industry-software/automation-software/scada/simatic-wincc-v7.html}}.
\bibitem[{Siemens(2020c)}]{Siemens2020}
\bibinfo{author}{Siemens}, \bibinfo{year}{2020}c.
\newblock \bibinfo{title}{{Totally Integrated Automation Portal}}.
\newblock \URLprefix
  \url{https://new.siemens.com/global/en/products/automation/industry-software/automation-software/tia-portal.html}.
\bibitem[{Siemens(2022)}]{SiemensHMI}
\bibinfo{author}{Siemens}, \bibinfo{year}{2022}.
\newblock \bibinfo{title}{Machine level hmi}.
\newblock
  \bibinfo{howpublished}{\url{https://new.siemens.com/global/en/products/automation/simatic-hmi/panels.html}}.
\newblock \bibinfo{note}{Last accessed: 21 March 2022}.
\bibitem[{Stouffer et~al.(2014)Stouffer, Pillitteri, Lightman, Abrams and
  Hahn}]{Stouffer2015}
\bibinfo{author}{Stouffer, K.}, \bibinfo{author}{Pillitteri, V.},
  \bibinfo{author}{Lightman, S.}, \bibinfo{author}{Abrams, M.},
  \bibinfo{author}{Hahn, A.}, \bibinfo{year}{2014}.
\newblock \bibinfo{title}{{Guide to Industrial Control Systems (ICS)
  Security}}.
\newblock \bibinfo{type}{Technical Report}. NIST.
\newblock \URLprefix
  \url{https://nvlpubs.nist.gov/nistpubs/SpecialPublications/NIST.SP.800-82r2.pdf}.
\bibitem[{Tamboli et~al.(2015)Tamboli, Rawale, Thoraiet and
  Agashe}]{tamboli2015implementation}
\bibinfo{author}{Tamboli, S.}, \bibinfo{author}{Rawale, M.},
  \bibinfo{author}{Thoraiet, R.}, \bibinfo{author}{Agashe, S.},
  \bibinfo{year}{2015}.
\newblock \bibinfo{title}{Implementation of modbus rtu and modbus tcp
  communication using siemens s7-1200 plc for batch process}, in:
  \bibinfo{booktitle}{2015 international conference on smart technologies and
  management for computing, communication, controls, energy and materials
  (ICSTM)}, \bibinfo{publisher}{IEEE}, \bibinfo{address}{New Jersey}. pp.
  \bibinfo{pages}{258--263}.
\bibitem[{Tenable(2022)}]{nessus}
\bibinfo{author}{Tenable}, \bibinfo{year}{2022}.
\newblock \bibinfo{title}{{Nessus Vulnerability Assessment}}.
\newblock
  \bibinfo{howpublished}{\url{https://www.tenable.com/products/nessus}}.
\bibitem[{Wardak et~al.(2016)Wardak, Zhioua and Almulhem}]{wardak2016plc}
\bibinfo{author}{Wardak, H.}, \bibinfo{author}{Zhioua, S.},
  \bibinfo{author}{Almulhem, A.}, \bibinfo{year}{2016}.
\newblock \bibinfo{title}{Plc access control: a security analysis}, in:
  \bibinfo{booktitle}{2016 World Congress on Industrial Control Systems
  Security (WCICSS)}, \bibinfo{publisher}{IEEE}, \bibinfo{address}{New Jersey}.
  pp. \bibinfo{pages}{1--6}.

\end{thebibliography}

\bio{}
Sam Maesschalck is a PhD student within the Lancaster University School of Computing and Communications and the Security Lancaster Institute. His research involves both technical (systems and networks) and non-technical (cyberspace and international relations) aspects of cyber security, mainly focused within a critical infrastructure environment. Currently his main focus is on the use on honeypots within ICS environments.
\endbio

\bio{}
Alexander Staves is a PhD student at the Lancaster University School of Computing and Communications and the Security Lancaster Institute. His research interests involve both penetration testing and cyber incident response and recovery for ICS/OT, in particular within critical infrastructure environments.
\endbio

\bio{}
Richard Derbyshire is a Senior Security Researcher at Orange Cyberdefense and an Honorary Researcher at Lancaster University, UK, where he obtained his PhD in computer science. His research involves both offensive and defensive elements of cyber security with a focus on penetration testing, risk assessment, and operational technology.
\endbio

\bio{}
Benjamin Green is an academic fellow in the School of Computing and Communications at Lancaster University, UK. His research involves both offensive and defensive elements of Industrial Control System security. He is involved in several related research projects, including Operational Technology Management after Cyber Incident (OT-MCI).
\endbio

\bio{}
David Hutchison is Emeritus Professor of Computing at Lancaster University, UK, and the Founding Director of InfoLab21. His work is well known internationally for contributions in a range of areas including Quality of Service, active and programmable networking, content distribution networks, and testbed activities. His current research focuses on the resilience of networked computer systems, and the protection of critical infrastructures and services.
\endbio

\appendix

\onecolumn
\section{Evaluated Library Functions}
\label{Appendix}

\begin{table}
\centering
\begin{tabular}{l l} 
\hline
\textbf{Function} & \textbf{Description}\\
\hline
\textbf{MODBUSPN} & Enables communication between a CPU\\
 & and a device which supports the Modbus/TCP protocol\\ 
\textbf{Automation System Mail (AS\_MAIL)} & Sends an e-mail from the CPU via a mail server \\
\textbf{Count Up (CTU)} & Increments the value at a given output \\ 
\textbf{Count Down (CTD)} & Decrements the value at a given output \\ 
\textbf{Generate Time Pulse (TP)} & Outputs a pulse for a programmed period \\
\textbf{Write Data Record (WRREC)} & Writes a data record to the addressed module \\
\textbf{Send Data (TSEND)} & Sends data over an existing communication connection \\
\textbf{Remote Write (PUT)} & Writes data to a remote CPU \\
\textbf{Communication Connection (TCON)} & Establishes a connection of type TCP or UDP from the user program \\ 
\textbf{Configure Communication Port (PORT\_CONFIG)} & change port parameters (such as data transmission rate) in runtime\\ 
\hline
\end{tabular}
\vspace{5pt}
\caption{Library Functions Tested with PLC-VBS (Siemens TIA v13)~\citep{Siemens2020}}
\label{tab:examplefunctions}
\end{table}

\end{document}